%% file: aa.tex
\documentclass[twocolumn]{aa} 

\usepackage[varg]{txfonts}
\usepackage{graphicx}
\usepackage{txfonts}
\usepackage{multirow}
\usepackage{xcolor}
\usepackage{tabularx}
\usepackage{array}
\usepackage{rotating}
\usepackage{makecell}
\usepackage{tablefootnote}
\usepackage{booktabs}
\usepackage{threeparttable}
\usepackage{siunitx}
\usepackage{soul}
\usepackage[utf8]{inputenc}
\usepackage{hyperref}
\def\ew{$W_{2796}$}

\def\ewfeii{$W_{2600}$}

\def\kms{km~s$^{-1}$}

\def\mgii{Mg~{\sc ii}~} 
\def\oii{O~{[\sc ii}]~} 
\def\oiii{O~{[\sc iii}]~} 
 
\def\mgiia{Mg~{\sc ii} $\lambda$2796} 
 
\def\mgiiab{Mg~{\sc ii} $\lambda\lambda$2796,2803~} 
\def\feii{Fe~{\sc ii}~} 
 
\def\feiia{Fe~{\sc ii} $\lambda$2600~} 
\def\oiiiab{[O~{\sc iii}] $\lambda\lambda$4959,5007} 
 
\def\oiiib{[O~{\sc iii}] $\lambda$5008~}

\def\oiii{[O~{\sc iii}]} 
\def\oiiab{[O~{\sc ii}] $\lambda\lambda$3727,3729}  
  
\def\oii{[O~{\sc ii}]}

\def\o3o2{[O~{\sc iii}]/[O~{\sc ii}]}

\def\hbeta{H$\beta$~}

\DeclareUnicodeCharacter{02BC}{'}
\input{numbers}
\begin{document}

   \title{Baryonic Ecosystem in Galaxies (BEINGMgII) - II. Unveiling the Nature of Galaxies Harboring Cool Gas Reservoirs
}


   \author{ 
        S. Das\inst{1}, 
        R. Joshi\inst{1},
        R. Chaudhary\inst{1},
        M. Fumagalli\inst{2,3},
        M. Fossati\inst{2,4}, 
        C. P\'eroux\inst{5,6},
        \and
        Luis C. Ho\inst{7,8}
    }

   \institute{Indian Institute of Astrophysics (IIA), Koramangala, Bangalore 560034, India \\     \email{rvjoshirv@gmail.com}
   \and
             Universit\'a degli Studi di Milano-Bicocca, Dip. di Fisica G. Occhialini, Piazza della Scienza 3, 20126 Milano, Italy
             \and
             INAF - Osservatorio Astronomico di Trieste, via G.B. Tiepolo 11, I-34143 Trieste, Italy
             \and 
             INAF - Osservatorio Astronomico di Brera, Via Brera 28, 21021, Milano, Italy
             \and
             European Southern Observatory, Karl-Schwarzschildstrasse 2, D-85748 Garching bei Munchen, Germany
             \and 
             Aix Marseille Universit\'e, CNRS, LAM (Laboratoire d’Astrophysique de Marseille) UMR 7326, F-13388 Marseille, France
             \and 
             Kavli Institute for Astronomy and Astrophysics, Peking University, Beijing 100871, Peopleʼs Republic of China
             \and
             Department of Astronomy, School of Physics, Peking University, Beijing 100871, Peopleʼs Republic of China
             }
   \date{Received July 26, 2024; accepted July 26, 2024}

 \titlerunning{BEINGMgII - Nature of Galaxies Harboring Cool Gas Reservoirs}
  \authorrunning{Das et al.}
  \abstract
    {We search for the galaxies associated with the intervening \mgii absorbers over a redshift range of $\totaldetectionzabsmin \le z \le \totaldetectionzabsmax$ using imaging data from DESI Legacy Imaging Surveys and measure the redshift based on direct detection of nebular emission in the background quasar spectra from SDSS survey.  We find \totaldetection\ \mgii absorbers associated with strong \oii\ nebular emissions, at $2.5\sigma$ level. Among them, for \primarydetection\ \mgii absorbers, we detect an absorber host galaxy at impact parameters of $\detectioniplowkpc \le \rho \le \detectioniphighkpc$ kpc, including a galaxy pair associated with three absorbers, with best-fit galaxy SED model based on multi-passband photometric data from DESI Legacy Imaging surveys, supplemented with the infrared VISTA and unWISE imaging surveys. The detection rate of \mgii absorber host with strong \oii\ nebular emission in the finite SDSS fiber of 2--3 arcsec diameter increases from 0.2\% to $\sim$3\% with increasing equivalent width from $0.3$\AA\ to $\sim$ 3.5\AA, which remains near-constant across the probed redshift range. 
    The associated \mgii host galaxies exhibit a wide range of stellar mass from \stellarmassmin $\le \rm log(M_{\star}/M_{\odot}) \le $\stellarmassmax, with an average star-formation rate (SFR) of $\rm\sfgsfrmedian M_{\odot}\ yr^{-1}$.  The \mgii absorber hosts selected based on \oii\ nebular emission mostly exhibit active star-forming systems including \sfrstarburstpercentage\% starburst systems, but \sfrpassivepercentage\% with suppressed SFR. The near-constant absorption strength at low-impact parameters suggests a high gas covering fraction. We find that the \mgii equivalent width (\ew) positively correlates with SFR and specific-SFR, likely indicate their wind origin. The average velocity offset between the host and absorber suggests that the \mgii gas is bound within the dark matter halo. }

  \maketitle
%

\section{Introduction}
\label{sec:intro}

The interplay between the galactic outflows and gas accretion from the diffuse gas reservoirs surrounding galaxies, i.e., circumgalactic medium (CGM) provides the key insight into the chemical enrichment history of galaxies \citep{peroux2020predictions}.  The quasar absorption line study has been proven to be a powerful tool for tracing the low-density gas in the CGM.  In particular, \mgiiab absorption line is a widely used tracer of cool  ($T \sim 10^4 K $) metal-enriched gas flows in and around galaxies \citep{Tumlinson2017ARA&A..55..389T}. Deep imaging and spectroscopic follow-up studies have revealed that strong \mgii\ absorbers (\ew\ $\ge$ 1\ \AA) are often associated with active phases of galactic evolution, such as star formation, starburst or AGN activity - originate in galactic disks on scales about $ 10$ kpc and in star formation-driven outflows on scales of $\gtrsim$ 100 kpc  \citep{Zibetti2007ApJ...658..161Z,Bouche2012MNRAS.426..801B, Kacprzak2014ApJ...792L..12K, Dutta2020MNRAS.499.5022D, Lundgren2021ApJ...913...50L}. The observed dependence of \mgii absorption on azimuthal angle, with a strong absorption above and along the disk plane, advocate bipolar outflows along the minor axis and inflows along the major axis \citep{Bouche2007ApJ...669L...5B, Nestor2011MNRAS.412.1559N, Lan2018ApJ...866...36L, Zabl2021MNRAS.507.4294Z, Guo2023Natur.624...53G, Bacon2023A&A...670A...4B} which is further supported by the simulations \citep{Peroux2020MNRAS.499.2462P, nelson2020resolving}. The majority of absorption seems having co-directional and co-planar accretion kinematics \citep{Kacprzak2011ApJ...733..105K,Lopez2020MNRAS.491.4442L,Nateghi2024MNRAS.534..930N}. \citet{Lan2018ApJ...866...36L} have shown that metal absorption is 5–10 times stronger around emission line galaxies on smaller scales than their red counterparts. However, the origin of \mgii absorption and the true nature of absorber host galaxies remains contentious.

Numerous efforts to understand the nature of absorber host galaxies have resulted in a few hundred galaxy–absorber pairs. These galaxies are generally found to be luminous (sub-L* type) at large impact parameters ranging from 10-200 kpc \citep{Chen2010ApJ...714.1521C,zabl2019muse, Dutta2020MNRAS.499.5022D}. Recent studies based on advanced integral field spectrographs show that the majority of \mgii absorbers are commonly associated with multiple galaxies which indicates that besides gas accretion and/or outflows the \mgii absorbers likely trace the multiple halos of a galaxy group \citep{Kacprzak2010MNRAS.406..445K,Bielby2017MNRAS.468.1373B, Peroux2017MNRAS.464.2053P,Peroux2019MNRAS.485.1595P,Dutta2020MNRAS.499.5022D} and/or intragroup medium \citep{Gauthier2013MNRAS.432.1444G}.

Note that finding the potential faint counterpart responsible for the absorption at small impact parameters has largely remained challenging, both due to the glare of bright background quasar and the faint nature of galaxy at high redshift \citep{Noterdaeme2010MNRAS.403..906N, Joshi2017MNRAS.471.1910J}. A handful of efforts, using high spatial resolution imaging in conjunction with multi-object spectroscopy, have bolstered the confidence of the presence of a faint but high star formation surface density absorber host at small impact parameters (\citealt{Bouche2007ApJ...669L...5B, Bouche2012MNRAS.426..801B, Lundgren2012ApJ...760...49L, Lundgren2021ApJ...913...50L}, see also, \citealt{Straka2015MNRAS.447.3856S,Joshi2017MNRAS.471.1910J,  Guha2022MNRAS.513.3836G, Guha2024MNRAS.527.5075G}). In our recent effort to probe the ultra-strong \mgii absorber in a deep imaging Hyper Supreme-Cam Subaru Strategic Program (HSC-SSP) Survey, we found a high detection rate of $\sim$\usmgiidetectionpercent\ percent at an impact parameter of $\le 30 \rm kpc$ (Joshi et al. 2024, submitted, hereinafter Paper I). The study of the physical origins of gas in the circumgalactic medium in TNG50 cosmological hydrodynamical simulation shows a probability of more than 50 percent for an absorber at small impact parameters of $\lesssim 20$ kpc being associated with a central galaxy of mass, log$M_{\star} \sim {8} M_{\odot}$ within
a typical $\Delta v$ range of $\sim$ 70 to 250 \kms \citep{weng2024physical}. However, the association of gas with absorbers in simulations over velocity space can select for gas beyond several times the virial radius \citep{rahmati2015distribution, ho2021identifying}. \par

To gain insights into the nature of absorber host galaxies and the potential origins of \mgii absorbers, in this paper, we extend the study from Paper-I to cover \ew\ range of $\gtrsim 0.3$\AA. In Paper I, we focus on the galaxies hosting ultra-strong (\ew $\ge 3$\AA) \mgii\ absorbers, primarily identified based on multi-band deep optical ($g, r, i, z, y$) HSC-SSP Subaru survey and near-infrared ($J, H, K$) VISTA survey SED model fitting, where 40\% of systems were spectroscopically identified from SDSS spectroscopic survey (see below).  In this study, we leverage the large sky coverage of $\approx$ 14,000 square degrees from the Dark Energy Spectroscopic Instrument (DESI) Legacy Imaging Surveys (DECaLS) to identify potential host galaxies. Given the limited passbands of DECaLS for obtaining reliable photometric redshifts, we focus exclusively on galaxies with robust spectroscopic redshift estimates derived from nebular emission lines observed in quasar spectra from the SDSS spectroscopic survey. \par

This paper is structured as follows. Section 2 describes our sample selection. This is followed by the methodology and analysis included in section 3. Section 4 examines the absorber galaxy association and the nature of galaxies associated with \mgii absorbers. The discussion and conclusions of this study are summarized in Section 5. Throughout, we have assumed the flat Universe with $H_0$ = 70 $\rm km\ s^{-1}\ Mpc^{-1}$, $\Omega_m$ = 0.3 and $\Omega_\Lambda$= 0.7.

\begin{figure*}
    \centering
    \includegraphics[width=0.95\textwidth]{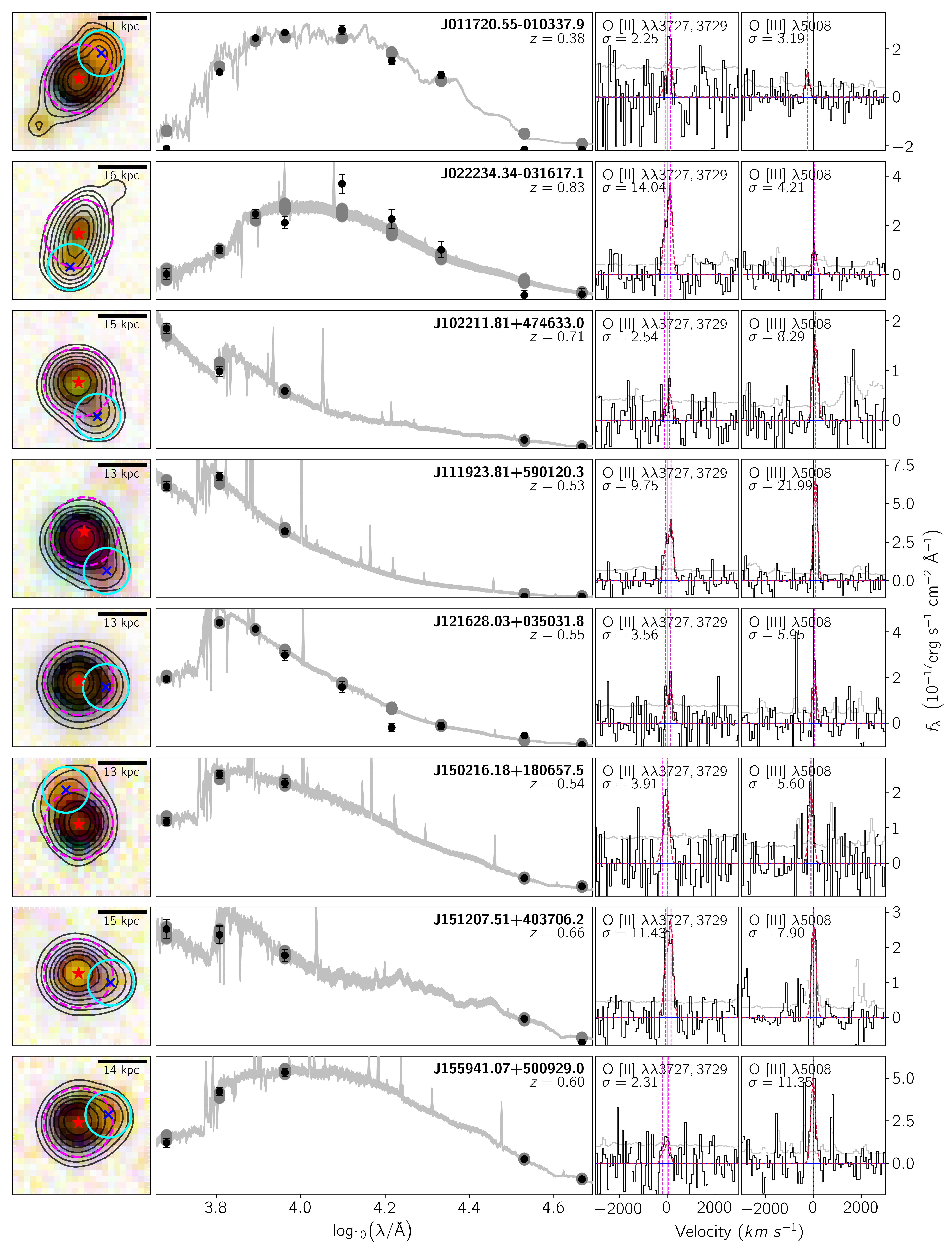}
    \caption{ The postage stamp DECaLS color composite images, centered on the quasar. The SDSS fiber, with a radius of \SI{1.5}{\arcsecond}, is indicated by dashed circles, while the cyan-colored aperture highlights the MgII host galaxy. The second column exhibits the multi-band SED fit at the absorber redshift. Columns three and four show the detected  \oiiab\ and \oiiib\ nebular emission line from \mgii absorber host, respectively.}
    \label{fig:collage}
\end{figure*}

\section{Data and Sample Selection}
Utilizing the \mgii absorber catalog from SDSS-DR16 \citet{Anand2022MNRAS.513.3210A}  and SDSS-DR7/DR12 \citet{Zhu2013ApJ...770..130Z} we assembled \numqsosightlinesoverall\ unique quasar sightlines tracing \numuniqueabsorbersoverall\ unique absorbers considering pairs of absorbers within 500 \kms\ as identical systems. Following Paper-I, we extend our search to all the \ew\ ranges by exploiting the imaging data from the DECaLS, spanning a wide sky coverage of $\approx 14000$ sq. degrees \citep{Dey2019AJ....157..168D}. The DECaLS offers multi-band imaging data in three optical passbands ($g, r, z$) at a $5\sigma$ survey depth of $g=24.0$, $r=23.4$, and $z=22.5$ AB magnitudes with a resolution of $\approx 0.262\si{\arcsecond}$ per pixel. The DESI footprint covers a total of \desicoveredqsosightlines\ quasar sightlines hosting \desicoveredabsorbers\ intervening absorbers within a broad redshift range up to $z\lesssim \desicoveredabsorberszmax$.  Considering the limitations in obtaining precise photometric redshifts with the limited passbands in the DECaLS survey, in this study we solely focus on spectroscopically identified galaxies. Therefore, for associating the absorber with galaxies, we restrict our search to sightlines with at least one photometric counterpart detected within 2 arcsec around the quasar. That is primarily to take advantage of the finite fiber diameter of 3\arcsec(2\arcsec) used in SDSS(BOSS) which integrates the light from all the objects falling in the fiber \citep{Noterdaeme2010MNRAS.403..906N, Joshi2017MNRAS.471.1910J}, and enables to directly detect the nebular emission (e.g., \oii, H$\beta$, O[{\sc~iii}]) from absorber host galaxies in the quasar spectra. Given the median seeing of $1.3\arcsec$ for DESI surveys, over the search radius of $2\arcsec$ centered on the quasar, a significant part of the galaxy may fall inside the fiber. This led to a sample of \desitwoarcsecdetabsorbers\ absorbers along \desitwoarcsecdetqso\ quasar sightlines with a total of \desitwoarcsecdetapertures\ galaxies detected within the defined search radius. Further, to avoid the false positive detection of nebular emissions due to poor sky removal, we avoid the wavelength region above $7500\AA$ where the \oiiab\ emission line falls in the sky region, translates to a redshift of $z \le 1$. This resulted in the final sample of \desifinalsampleabsorbers\ absorbers along \desifinalsampleqso\ sightlines with \desifinalsampleapertures\ galaxies detected within the defined search radius.

\begin{figure*}
    \centering
    \includegraphics[width=0.89\textwidth]{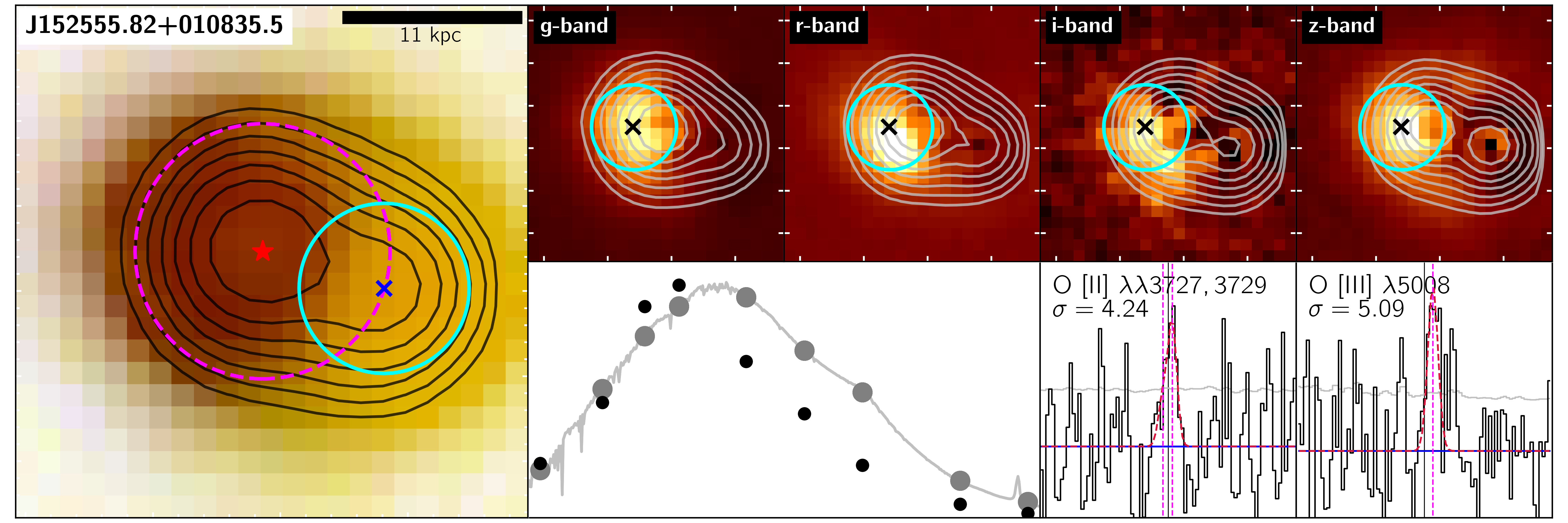}
    \caption{ {\it Left Panel:} The multi-band color composite image centered on quasar J152555.82+010835.5. The quasar and absorber host galaxy candidate detected in DECaLS survey within 2 arcsec are marked as {\it star} and {\it cross} symbols. In addition, the dashed and solid circle shows the SDSS fiber of radius 1.5 arcsec and absorber candidate. The contour represents the best-fit model for quasar and galaxy provided by DECaLS. {\it Right upper panels:} show the residual $g, r, i, z$ passband images. The {\it cross} symbol shows an additional stellar counterpart not detected by DECaLS, but detected in the residual images of all the passbands, which may likely be the potential host. {\it Lower panels:} shows the SED fit at the absorber redshift of $z_{abs}=0.39$ along with the \oiiab\ and \oiiib\ nebular emission.}
    \label{fig:residual}
\end{figure*}

\section{Analysis}

\subsection{Searching for the nebular emissions}

We search for the \oiiab\  emission line for each \mgii absorber within the velocity range of $\pm 400 \rm km s^{-1}$ around the absorber redshift ($z_{abs}$) in the quasar continuum subtracted spectrum. The detection significance of the emission line feature is determined based on the signal-to-noise ratio defined in \citet{Bolton2004AJ....127.1860B} as $S/N = \frac{\sum_{i}^{} f_i u_i / \sigma_i^2}{\sqrt{\sum_{i}^{} u_i^2/\sigma_i^2}}$, where $f_i$ is the residual line flux in the $i^{th}$ pixel, $\sigma_i$ is the flux error and $u_i$ is a Gaussian kernel, normalized such that $\sum_i u_i = 1$. The position and line width of the kernel are determined by minimizing the $\chi^2$ over the defined velocity window.

In addition to the \oii\ emission line, we also search for the H$\beta$, and \oiiiab\ nebular emissions. Given that all our quasar sightlines host at least one close companion, we considered any feature detected at $\ge 2.5\sigma$ as positive \oii\ detection, resulting in \oiitwopfivedetection\ absorbers which include \oiithreedetection\ systems with \oii\ detection at $\ge 3\sigma$ level. Alongside, we also consider \oiiihbadditionaldetection\ systems with the \oiiiab\ and/or H$\beta$ nebular emission detected at $\ge 3\sigma$ level. In addition, we performed visual scrutiny for any bad pixel, continuum, or strong skyline regions to avoid the false positive, resulting in total \totaldetection\ \mgii absorbers with associated nebular emissions. Of which, 242 are new detections and 28 absorbers are part of the galaxy on top of quasar sample from \citet{Joshi2017MNRAS.465..701J}.

\subsection{Origin of nebular emission: absorber--galaxy association}
For the absorber-galaxy association of the candidate absorber host galaxy detected at low \oii\ SNR of 2.5$\sigma$, following paper-I, we further performed the multi-band spectral energy distribution fitting to obtain robust photometric redshift.  The DECaLS optical passband, $g, r, i, z$, fluxes were supplemented with the forced photometry of optically detected sources in mid-infrared passbands $W1$, and $W2$ from unblurred Wide-field Infrared Survey Explorer (unWISE) coadded images reaching up to $5\sigma$ depth of $20.0$ and $19.3$ AB magnitudes in $W1$ and $W2$ respectively  \citep{Dey2019AJ....157..168D}. The DECaLS fluxes are obtained by modeling each source, detected in three individual-band image stacks at 6$\sigma$ level, using the Tractor algorithm which works on pixel-level data. In brief, the Tractor models each source using a small set of parametric light profiles, including a delta function for point sources, a de Vaucouleurs law, an exponential disk, or a composite de Vaucouleurs plus exponential. The best-fit model is determined by convolving each model with the specific PSF for each exposure, fitting to each image, and minimizing the residuals for all images.  Assuming that the model is the same across all the bands, e.g,  if the source is spatially extended, then the same light profile (an exponential disk, de Vaucouleurs, or combination) is consistently fit to all images to determine the best-fit source position, source shape parameters, and photometry \citep{Dey2019AJ....157..168D}.  To extract the mid-infrared photometry of W1, and W2 WISE bands, having a resolution of 6 arcsecs, forced photometry is performed by modeling each optically-detected source forcing the location and shape of the model by convolving with the WISE PSF and fitting to the WISE stacked image \citep[see also,][]{Lang2016AJ....151...36L}.

In addition, we also search for near-infrared, $J, H, Ks$, passband images from the VISTA Surveys\footnote{\href{http://vsa.roe.ac.uk/}{http://vsa.roe.ac.uk/}}. We obtained \vhsdetection\ source in the VHS (at 5$\sigma$ depth of J= 21.2, H=20.6, Ks=20.0) survey and \vikingdetection\ in VIKING (at 5$\sigma$ depth of J= 22.1) survey. For the near-IR flux measurement, we simultaneously model the quasar and the galaxy at the centroid obtained from DESI images. In the case of non-detection, we have estimated flux upper limits at 3$\sigma$ level. To obtain more secure results, we demand that the target be detected (at $3\sigma$ level) in any of the three filters mentioned above, resulting in \threefilterdetection\ sources, of which more than $\sim$75\% of the systems having minimum 4-band detection.

Next, we model the galaxy SED at fixed absorber redshift using {\sc BAGPIPES}\footnote{\href{https://bagpipes.readthedocs.io/en/latest/}{https://bagpipes.readthedocs.io/en/latest/}} \citep{Carnall2018} utilizing the MultiNest sampling algorithm \citep{Feroz2009}. We use a simple model considering a delayed star formation history with a wide parameter space for age between 50 Myr to 13.5 Gyr, mass formed ( $ 6 \le \rm log(M_*/M_{\odot}) \le 13$), and metallicity ($\rm 0.005 < [Z/H] < 5$). We assume the dust extinction law of \citet{Calzetti1994ApJ...429..582C} with total extinction $0 < A_v < 4$. \par

For the \totaldetection\ \mgii absorbers with \oii\ detection $\ge 2.5\sigma$ level or other emission lines (i.e, \hbeta, \oiii),  we found \totalhostdetection\ sources within the search radius of $2\arcsec$. Out of these sources, we obtained \threefilterdetectiongoodsed\ bonafide sources with good SED fit (see Figure~\ref{fig:collage}). The remaining \noandbadsed\ sources with robust \oii\ detection, but poor SED fit either due to limited filters or the fact that absorption is produced by some other undetected faint nearby galaxy, we analyzed DECaLS residual images for all the sources to detect any possible galaxy counterpart. For 9 sources, we detect a non-zero flux residual at $> 3\sigma$ level. Figure~\ref{fig:residual} shows an example of an absorber galaxy candidate within 2 arcsecs around a quasar with strong \oiiab\ \ and \oiiib\ nebular emission, but poor SED fit at the absorber redshift. The residual images reveal an additional stellar counterpart detected at $\ge  3 \sigma$ in all the passbands, marked as a cross symbol, which may likely be the true absorber host. Hereafter, to study host galaxy properties of \mgii absorbers, we only consider \primaryhostdetection\ objects with good SED fit. The remaining \secondarydetection\ systems are considered bonafide cases of galaxy on top of quasar and included as part of the online catalog as supplementary material. The properties of the large set of \mgii absorber host galaxies without detected nebular emission lines, but good SED fits are further explored in Chaudhary et al. (in preparation). \par

\begin{figure*}
    \centering
    \includegraphics[width=0.48\textwidth]{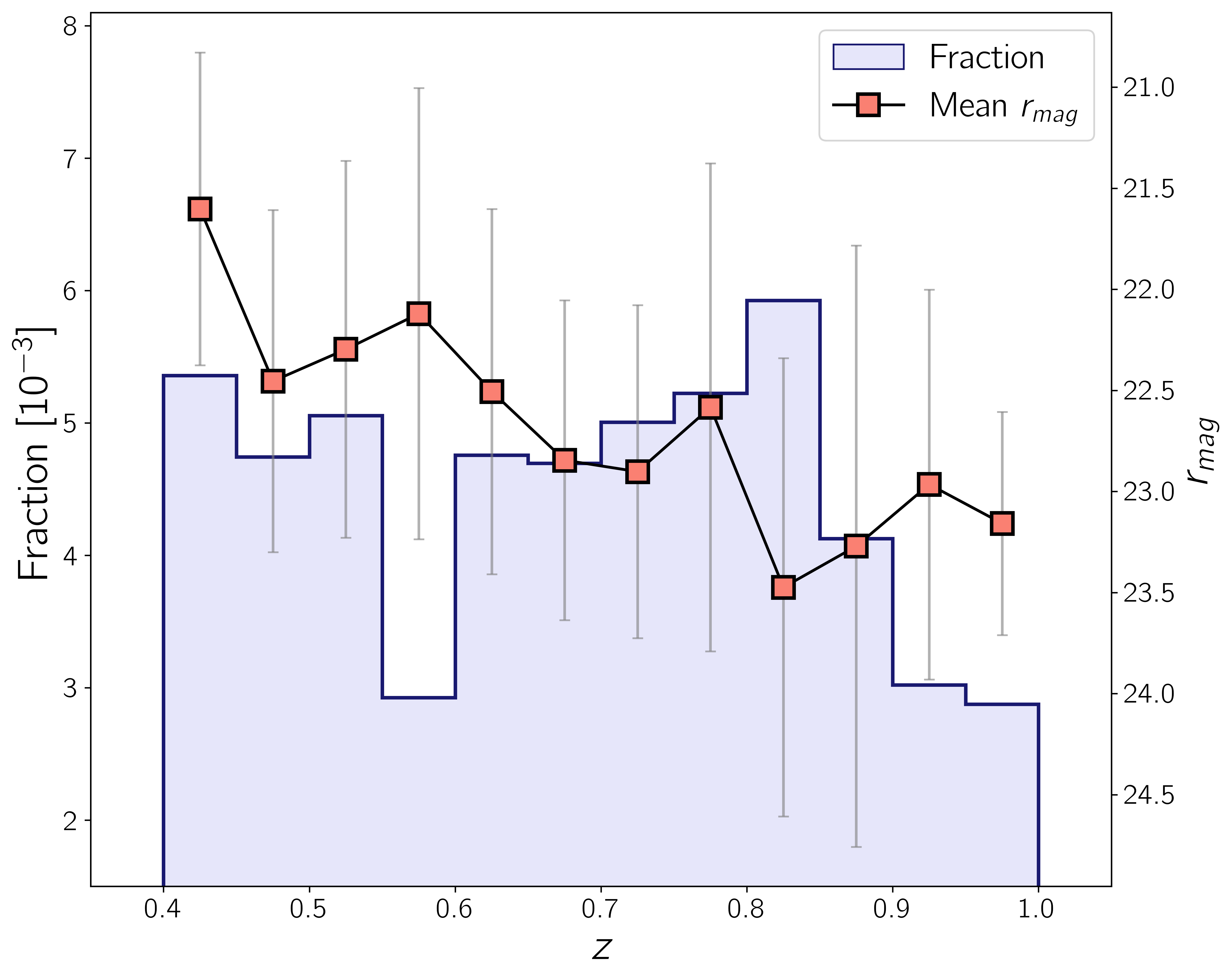}
    \includegraphics[width=0.48\textwidth]{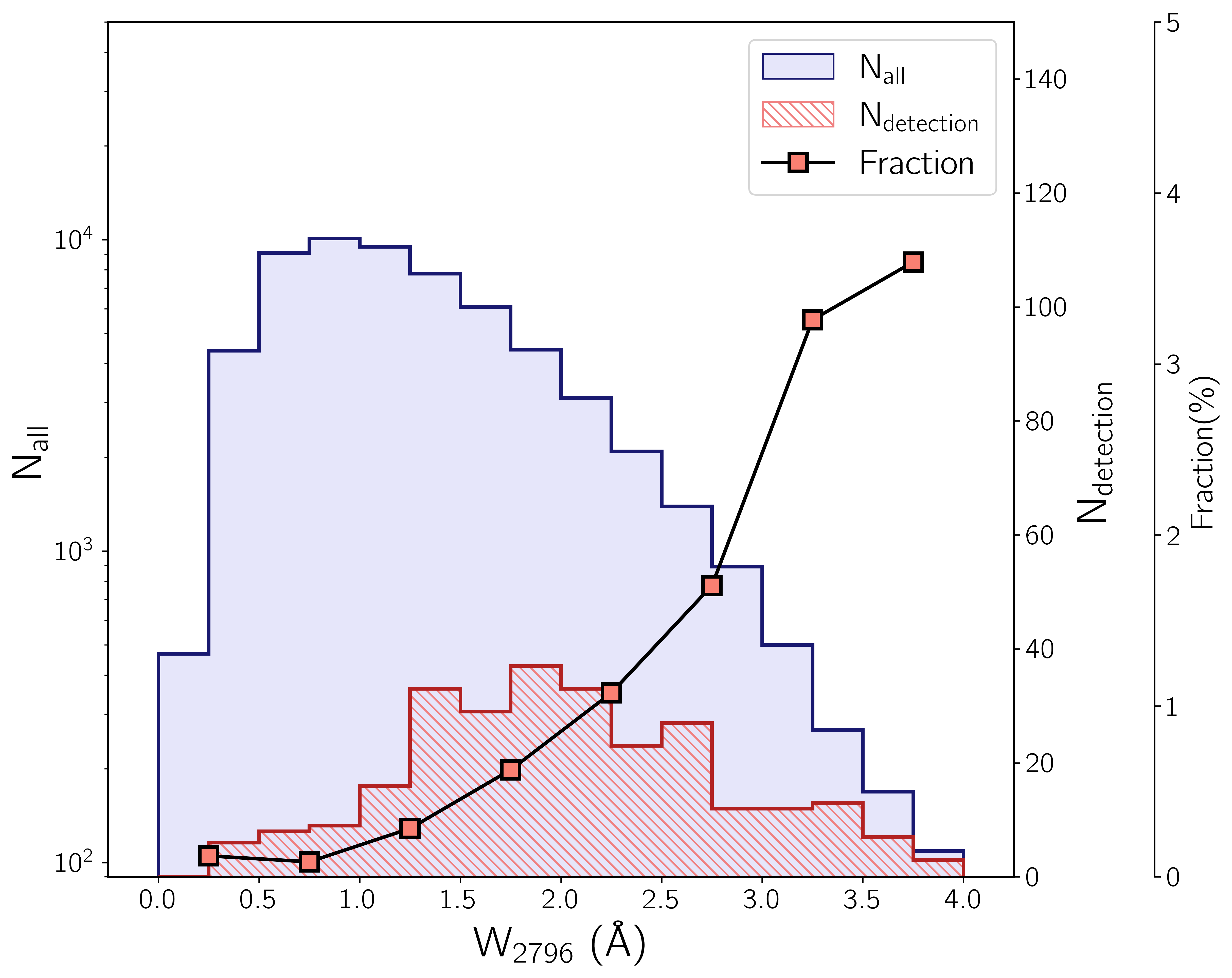}
    \caption{ {\it Left Panel:} The detection fraction of \mgii absorber as a function of redshift. The average DECaLS $r_{mag}$ of galaxies per redshift bin is shown as {\it square}, along with the 16th and 84th percentiles. {\it Right Panel:} Equivalent width (\ew) distribution of \mgii absorbers for entire sample   with $z < 1$ ({\it solid histogram}). The ({\it hatched histogram}) shows the distribution for the detection set, given in the right-hand side ordinates. The detection fraction per \ew\ bin is shown as {\it square} symbol given in the additional y-axis, the right-hand side ordinates.}
    \label{fig:detection}
\end{figure*}

\section{Results}

In Figure~\ref{fig:collage}, we show a few examples of \mgii absorber galaxies. Given the quasar proximity, i.e., within 2\arcsec, enabled us to map the gas in galaxy halo or ISM over physical scales of $\sim \detectioniplowkpc-\detectioniphighkpc$ kpc over the redshift range of $\detectionzabsmin < z < \detectionzabsmax$. Here, we discuss the dependence of absorber properties on the host galaxies.

\subsection{Detection rate of emission selected \mgii absorber galaxies}

 In {\it right panel} of Figure~\ref{fig:detection}, we show the rest-frame equivalent width (\ew) distribution of our primary sample of  $z < 1$ \mgii\ absorbers along with the absorbers with direct detection of \oii\ nebular emission. The majority, \mgiiewgeonepercentage\%, of \mgii systems, belong to the strong \mgii absorbers with \ew\ $\ge 1$\AA, with a median \ew\ of \medianmgiiew\AA. The detection fraction of \mgii galaxies at proximity to the quasar increases with \ew, rising from $\sim 0.2$ percent for an absorber with \ew\ of $\sim$1\AA\ to more than  2 percent for \ew\  $ \ge 3 \AA$.  Furthermore, we compare the detection fraction of emission-selected galaxies as a function of redshift. The {left panel} of Figure~\ref{fig:detection} shows the distribution of detection fraction typically ranging between  0.006 to 0.003 from redshift 0.4 to 1,  with a median of 0.005. Note that, the detection fraction mentioned above represents a lower limit, constrained by the finite fiber size, and limited imaging and spectroscopy sensitivity in identifying faint galaxies. The above detection rate is consistent with the spectroscopically selected galaxy on top of quasar cases by \citet{Joshi2017MNRAS.471.1910J} which varies between $\sim 1-3$ percent for \mgii absorbers with \ew\ $> 2$\AA\ from SDSS-DR7 and SDSS-DR12, respectively. In contrast to the low detection rates observed in studies, including the present one, that rely on nebular emission line detection in quasar spectra with finite fiber size, in Paper-I, we find a significantly higher detection rate of around 40\% for ultra-strong \mgii\ absorbers. This highlights the importance of deep imaging and spectroscopic surveys to detect faint absorber hosts and study their metal-enriched halos.

\subsection{Star-formation rate of \mgii\ absorber host galaxies}

Using our sample of \mgii galaxies with direct detection of \oii\ and \oiii\ nebular emissions, we derive the SFR assuming stars as the ionizing source, and using prescription given by \citet{Kennicuttb1998ApJ...498..541K}, i.e., SFR\oii  = $(1.4 \pm 0.4) \times 10^{-41}L$\oii. The \oii\ SFR is found in the range of \sfroiimin\ to \sfroiimax\ $\rm M_{\odot}\ yr^{-1}$. Considering the \oii\ luminosity function of galaxies at $z = 0.65$ from \citet{Comparat2016MNRAS.461.1076C}, our measured \oii\ nebular line luminosity ranges between 0.14–3.5 $L^{\star}$\oii. We note the measured SFR are lower limits as we do not apply any correction for the dust reddening and the emission line fluxes are affected by fiber losses \citep{Lopez2012MNRAS.419.3553L,Joshi2017MNRAS.471.1910J}. \par 

\begin{figure}
    \centering
    \includegraphics[width=0.49\textwidth]{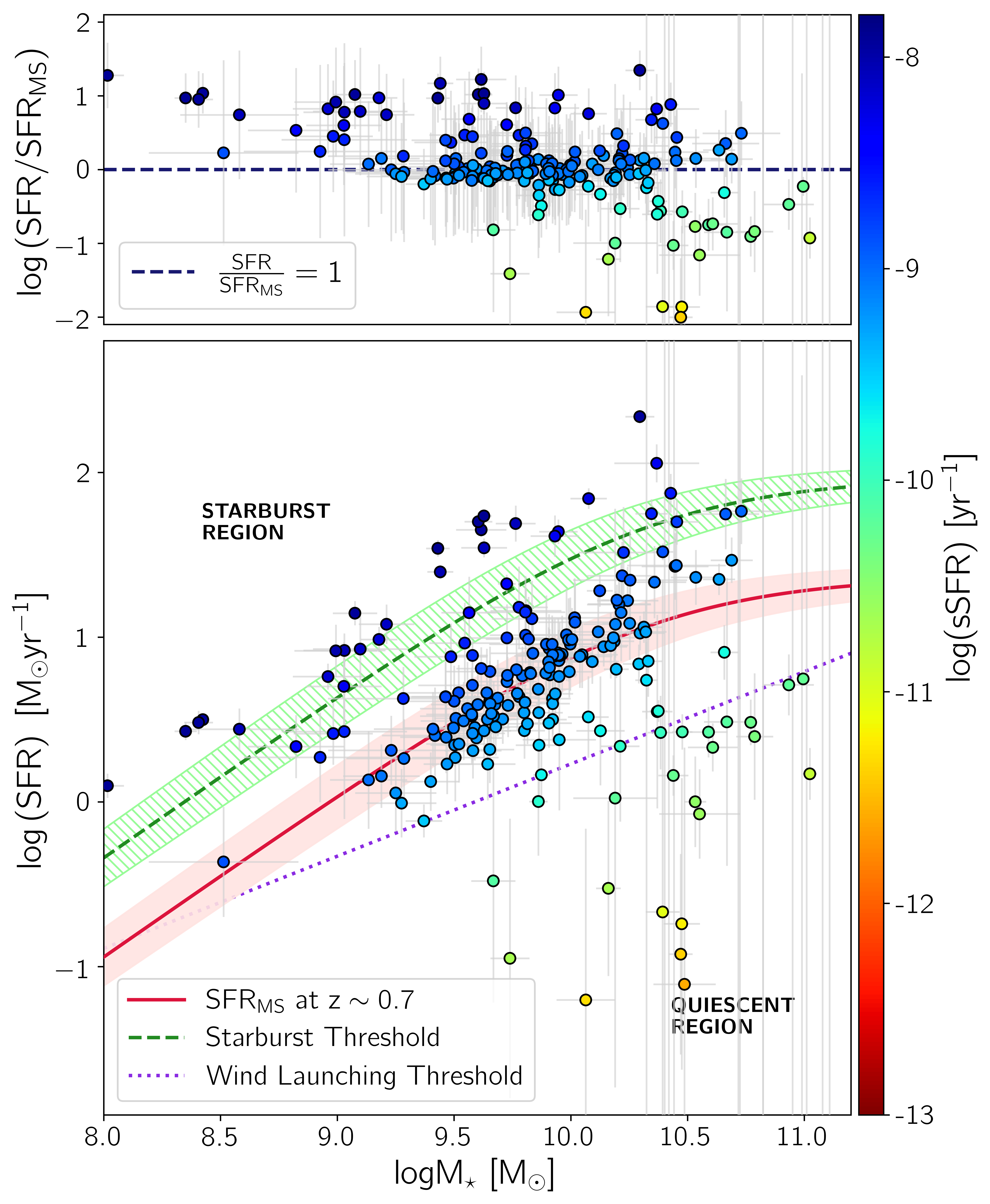}
    \caption{ Star formation Main sequence relation for MgII absorber host galaxies. The symbols are color-coded with a specific star-formation rate. The {\it shaded} and {\it hatched} regions indicate the $1\sigma$ confidence interval. Additionally, the {\it dotted-line} represents the SFR threshold for wind launching from \citet{Murray2011ApJ...735...66M}. {\it Top panel:} SFR of \mgii galaxies normalized by the main sequence star-forming galaxy at respective redshift.}
    \label{fig:SFR_SM}
\end{figure}

We derive the physical parameters, i.e., stellar mass, and star-formation rate, using {\sc bagpipes} multi-band SED modelling. The \mgii galaxies in our sample trace the stellar mass ranging between $\rm \stellarmassmin \le log (M_{\star}/  M_{\odot}) \le \stellarmassmax$, with an average $\rm \langle log\ M_{\star} \rangle$ of \stellarmassavg\ $\rm M_{\odot}$. Out of \primaryhostdetection\ galaxies, \detectionsfg\ are star-forming systems defined as $\rm \log{sSFR}>-10.6\ yr^{-1}$. These galaxies exhibits a typical star-formation rate of $\rm \sfgsfrmin \le  SFR [M_{\odot} yr^{-1}] \le \sfgsfrmax$, with a median  $\rm \langle SFR \rangle$ of \sfgsfrmedian\ $\rm M_{\odot}\ yr ^{-1}$. A handful, i.e., \sfrpassivepercentage\% of \mgii galaxies show passive nature with   $\rm \log{sSFR} < -10.6\ yr^{-1}$ and $\rm \langle SFR \rangle$ of \passivesfrmean\ $\rm M_{\odot}\ yr ^{-1}$. In Figure~\ref{fig:SFR_SM}, we compare the stellar mass versus star-formation of our \mgii absorber galaxies. For reference, we show the best-fit relation for main sequence star-forming galaxies at $\langle z \rangle \sim  0.7$  from \citet{Popesso2023MNRAS.519.1526P}.  As our galaxies are selected based on \oii\ nebular emission, it is evident that the majority of \mgii\ absorber hosts are star-forming galaxies, following the star formation main sequence. The same is clear from the top panel of Figure~\ref{fig:SFR_SM} showing the SFR, normalized by the main sequence star-formation at the respective stellar mass and redshift, using the same best-fit relation depicted in the bottom panel. About \sfrstarburstpercentage\% of the systems with SFR of factor 3 to 4 higher than the main sequence show a starburst nature \citep{Elbaz2018A&A...616A.110E}, whereas a large fraction has sufficiently higher star-formation rate to launch strong outflows \citep{Murray2011ApJ...735...66M} (see, Figure~\ref{fig:SFR_SM}).

\subsection{Dependence of \mgii absorbers on galaxy properties}

We explore the various correlations between the absorber properties such as \ew\ strength, equivalent width ratio between \feii and \mgii absorption (\ewfeii/\ew) that maps the $\alpha$ enrichment, and impact parameter ($\rho$) with the galaxy properties, such as M$_{\star}$, and SFR. \par 

\begin{figure*}
    \centering
    \includegraphics[width=0.9\textwidth]{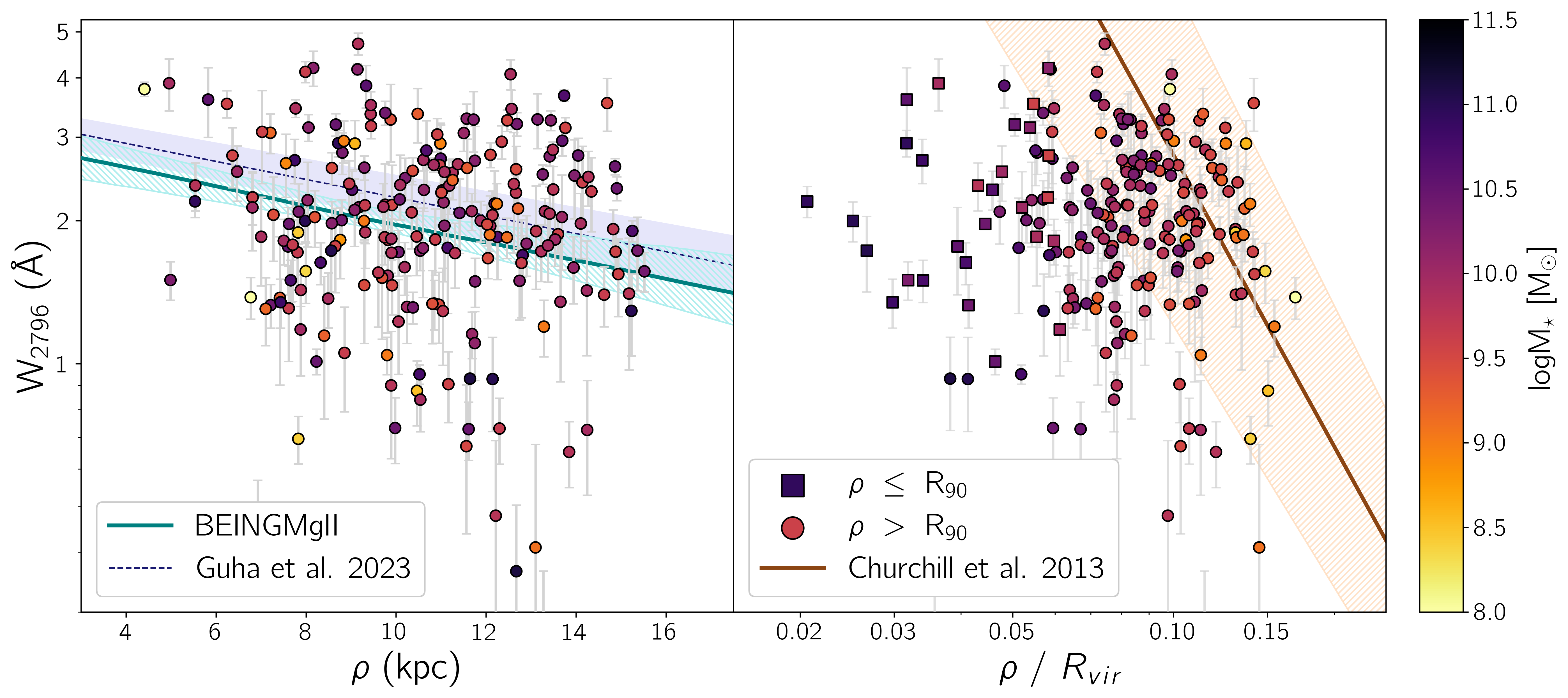}
    \caption{ {\it Left panel:} The \mgii absorber rest-frame equivalent width versus impact parameter along with best fit shown as hatched region. The \ew-vs-$\rho$ best-fit relation from \citet{Guha2022MNRAS.513.3836G} is shown as a shaded region. {it Right panel:} The \mgii absorber equivalent width versus impact parameter normalized by the galaxy's virial radius ($\rho/R_{vir}$). The best-fit relation from \citet{Churchill2013ApJ...763L..42C} is shown as a hatched region. The galaxies with impact pentameters below $R_{90}$ radii are shown as {\it square}. The symbols are color-coded with stellar mass.}
    \label{fig:EW_D}
\end{figure*}

Firstly, we test the firmly established anti-correlation between \ew\ and ($\rho$). The {\it left panel} of Figure~\ref{fig:EW_D} show the \ew\ as a function of $\rho$. To examine the dependence of \mgii absorption strength on galactocentric distance, we fit a log-linear model as, log\ew\ (\AA) = $\alpha$ + $\beta \times \rho$ (kpc), with a likelihood function given in equation 7 of \citet[][]{Chen2010ApJ...714.1521C}, and sample the posterior probability density function using PyMultiNest. For the  \mgii galaxies with lower $\rho$ in our sample we 
obtained a best-fit parameter of $ \alpha = 0.494^{+0.066}_{-0.062}, \ \beta = -0.02^{+0.006}_{-0.008}$. Figure~\ref{fig:EW_D} shows the best-fit log-linear model, along with 1$\sigma$ uncertainty in the shaded region. The \ew\ is mildly anti-correlation with $\rho$ which is consistent with the previous studies \citep{Chen2010ApJ...714.1521C, Nielsen2013ApJ...776..115N}. This hints that the absorbing gas possess a higher covering fraction at small impact parameters of $\le 20$kpc \citep[see also,][]{Dutta2020MNRAS.499.5022D}.  \par

\begin{figure}
    \centering
    \includegraphics[width=0.49\textwidth]{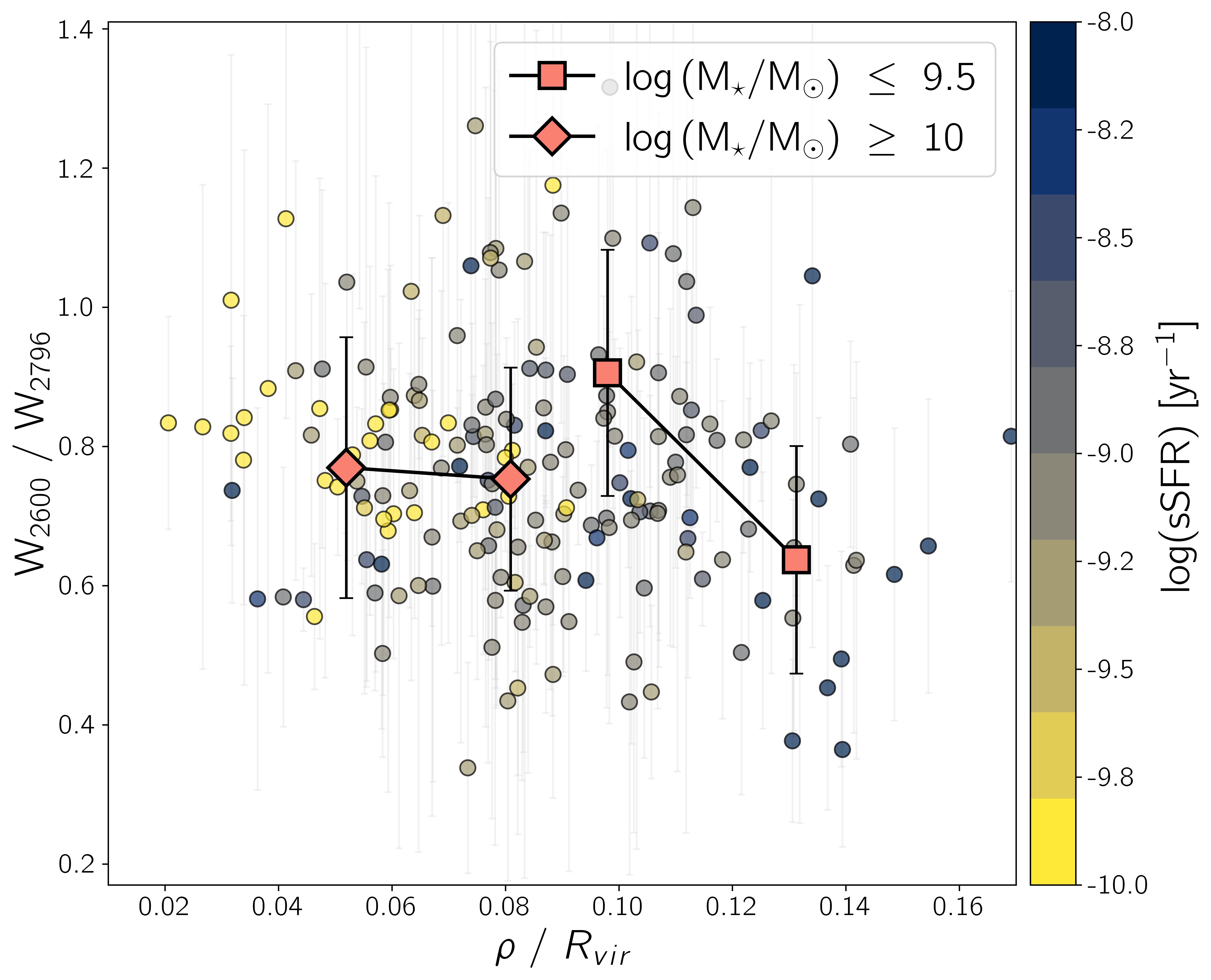}
    \caption{The rest-frame equivalent width ratio between  \feiia and \mgiia\ absorption versus impact parameter normalized by the galaxy's virial radius ($\rho/R_{vir}$). The symbols are color-coded with specific star-formation rates.}
    \label{fig:feii}
\end{figure}

Previous studies have revealed an empirical correlation between the  \ew\  and galaxy's stellar mass  \citep{Churchill2013ApJ...763L..42C, Dutta2020MNRAS.499.5022D}. For the $\rho$ range in our study, we find that the \ew\ for the massive galaxies ($\rm log M_{\star} \ge \langle \log{M_{\star}} \rangle$) are not significantly different compared to the low-mass galaxies based on a Kolmogorov-Smirnov (KS) test statistic of $D=0.05$ at $p_{null}=0.99$. We further investigated the influence of stellar mass on the \ew–$\rho$ anti-correlation for the above two subsets using a two-dimensional KS test. The result suggests that the distributions of the \ew–$\rho$ are likely drawn from the same parent population, with $D$ = 0.097 and $p_{null} = 0.83$. In an analysis of 183 isolated \mgii absorber galaxies over $0.07 \le z  \le 1.12$ within projected galactocentric distances $\rho \sim$ 200 kpc, \citet{Churchill2013ApJ...763L..42C} found that the scatter in \ew-$\rho$ plane vanishes when accounted for the stellar mass. Therefore, we compare the \ew\ with $\rho/R_{vir}$, where $R_{vir}$ is virial radius defined as an overdense region with a density 200 times the critical cosmic density $\rho_{crit}$, i.e. $R_{\rm vir} = (M_h/(3/4\pi) 200\ \rho_{crit})^{1/3}$. Here, we estimate the virial radius by converting the stellar mass into the halo mass using the stellar-to-halo mass relation from \citet{Girelli2020A&A...634A.135G}. The {\it right panel} of Figure~\ref{fig:EW_D} show \ew\ versus $\rho/R_{vir}$ along with the best relation from \citet{Churchill2013ApJ...763L..42C}. We note the large scatter in \ew\ distribution may result from the sightlines tracing the galaxy ISM. To test this possibility, using the universal stellar mass–size relation of galaxies from \citet{Ichikawa2012MNRAS.422.1014I}, we measure the 90 percent-light ($R_{90}$) radii of galaxies and compare the relation between \ew\ versus  $\rho/R_{vir}$ for the systems with $\rho \ge R_{90}$, likely tracing the gas in the halo and disk-halo interface (see Figure~\ref{fig:EW_D}, right panel). The evident tighter relation between \ew\ and $\rho/R_{vir}$ implies that the \mgii\ gas is strongly linked to the halo mass \citep[see also, ][]{Churchill2013ApJ...763L..42C}.

Considering that our detection probes low-impact parameters of $< 20$ kpc, it offers to probe the disk halo interface where the impact of galactic outflows is more pronounced.  The metal enrichment of the CGM is directly related to the supernova rate. Therefore,  the ratio of $\alpha$-elements to Fe should reflect the supernova rates and efficiency of outflows in enriching the CGM. \citet{Joshi2018MNRAS.476..210J} have shown that the \mgii\ absorbers detected in nebular emission in the SDSS fiber spectra of quasar show a strong \feii absorption with \ewfeii/\ew $> 0.5$. We detect a similar trend with strong emission line galaxies associated with higher \ewfeii/\ew\ ratio ranging between $0.5-0.8$. In Figure~\ref{fig:feii}, we compare the \ewfeii/\ew\ as a function of $\rho/R_{vir}$. The \ewfeii/\ew\  show no correlation with  $\rho/R_{vir}$ with the Kandall's $\tau$ coefficient of  $\tau=$\feiimgiivsrhorvirstat\ with $P_{null} =$\feiimgiivsrhorvirpvalue. Furthermore, to check the dependence of $\alpha$ enrichment on stellar mass,  we divide our sample into dwarf galaxies with log(M$\rm _{\star}$/M$_{\odot}) \le 9.5$ and massive galaxies with log(M$\rm _{\star}$/M$_{\odot}) \ge 10$. We compare the  \ewfeii/\ew\ for a subset of \dwarfgalaxydetection\ dwarf galaxy split at median $\rho/R_{vir}$, shown as {\it square} symbol in Figure~\ref{fig:feii}. The declining trend in the \ewfeii/\ew\ ratio with increasing distance is clearly evident and corroborated by Kendall's correlation coefficient of $\tau=-0.48$ at $p_{null}<0.01$. Interestingly, the subset of  massive galaxies with log(M$\rm _{\star}$/M$_{\odot}) \ge 10$ in our sample show   relatively lower \ewfeii/\ew\  at smaller  $\rho/R_{vir}$.  
The relatively lower $\rho/R_{vir}$ for the massive galaxies in our sample, preferentially probing the extended disk with higher dust, may likely give rise to lower $\alpha$ enrichment due to excess depletion of  \feii\  on to dust grains. To test this scenario, following \citet{Srianand2008MNRAS.391L..69S}, we measure the selective reddening $E(B - V)$ by fitting the quasar composite spectrum, reddened by the Small Magellanic Cloud (SMC) extinction curves \citep{Gordon2003ApJ...594..279G}. The median reddening along \mgii sightlines is found to be $E(B - V) = \ebvmedian$, which is similar to found for high H{\sc~i} columns density (N(H{\sc~i})$\ge 10^{20.3} \rm cm^{-1}$) damped Ly$\alpha$ absorber  systems \citep{Murphy2004MNRAS.354L..31M}. The higher \feii depletion for the above subset of massive galaxies is also supported by higher  $E(B - V) = \massiveebvmean \pm \massiveebverror$, than $E(B - V)=$ \dwarfebvmean $\pm$ \dwarfebverror\ for the dwarf galaxy subset.

\begin{figure*}
    \centering
    \includegraphics[width=0.95\textwidth]{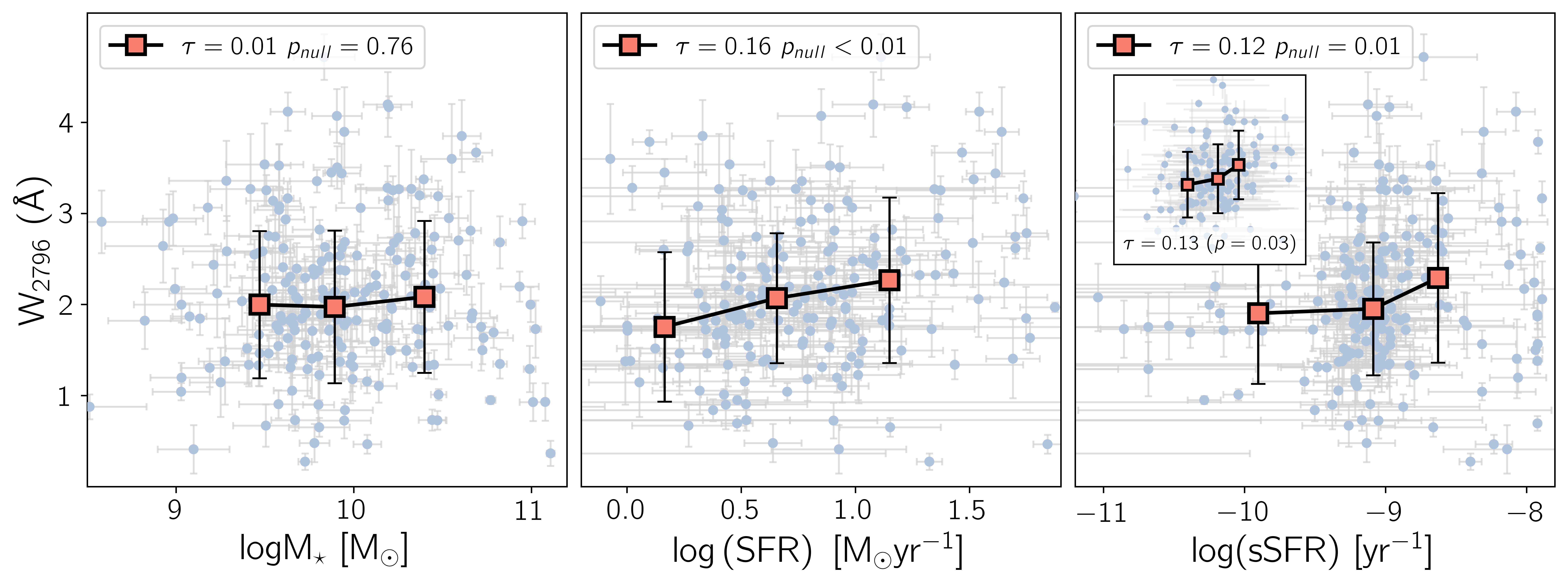}
    \caption{Distribution of \mgii equivalent width with the galaxy stellar mass, star formation rate, and specific star-formation rate. The square symbols represent the median values over three equal bins.}
    \label{fig:corr}
\end{figure*}

\begin{figure}
    \centering
    \includegraphics[width=0.49\textwidth]{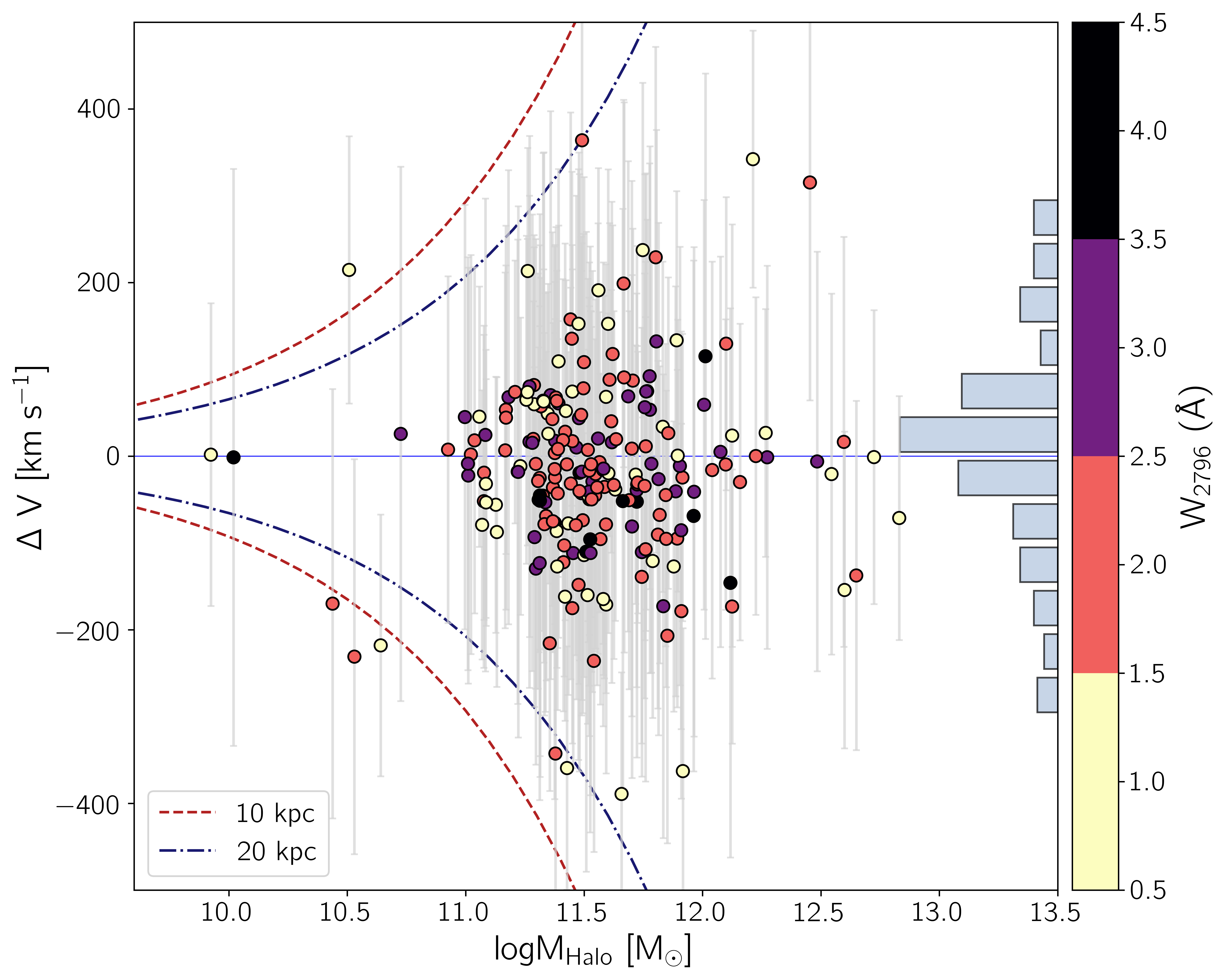}
    \caption{\mgii component velocity centroids with respect to the galaxies' systemic redshift as a function of the inferred dark matter halo mass. The histogram represents the distribution of the individual component velocities. The dashed and dashed dots show the mass-dependent escape velocities at a distance corresponding to the average and the maximum impact parameters of 10 and 20 kpc, respectively in our sample.}
    \label{fig:DELTAV_MHALO}
\end{figure}

In Figure~\ref{fig:corr}, we show the \ew\ dependence on $M_{\star}$, SFR, and specific SFR (sSFR). The median \ew\ over three equal bins is also shown as square. The strength of \ew\ is mildly correlated with the SFR and sSFR with Kendall's $\tau$ correlation coefficient of $\tau_k$= 0.16, 0.12 with a null probability of correlation arising by chance is $p_{null} \lesssim 0.01$, respectively.  This likely implies that the positive correlation between \ew\ versus sSFR is driven by SFR and not the M$_{\star}$. The inset in the right panel of Figure~\ref{fig:corr} shows the galaxies with SFR consistent with the main sequence star-forming galaxies at the respective redshift within the $1\sigma$ range. For this subset of main sequence galaxies, the \ew\ and sSFR positively correlate with Kendall's-$\tau$ correlation coefficient of $\tau= 0.13$, and $P_{null} = 0.03$.

\subsection{\mgii gas-galaxy kinematics}
Here, we test if the \mgii gas is kinematically bound to the halo by using the relative galaxy and absorption velocities and compare it with the escape velocities of their halos. Note that apart from the outflows, the velocity offset may result from the accreting material from the intergalactic medium or galactic fountains which is difficult to disentangle. Figure~\ref{fig:DELTAV_MHALO} shows escape velocities at an average impact parameter of 10 kpc maximum impact parameter of $\sim$ 20 kpc, respectively, in our sample, computed for spherically symmetric Navarro–Frenk–White dark matter halo profile \citep{Navarro1996ApJ...462..563N}, as a function of halo mass. The halo mass is derived based on the respective absorber redshifts and the stellar masses using the stellar-to-halo mass relation given by \citet{Girelli2020A&A...634A.135G}. It is evident that the majority of the absorption velocity centroids are below the estimated galaxy halo escape velocities. It is worth noting that measured relative velocity depends on the galaxy inclinations and quasar sight-line azimuthal angles. Therefore, it is likely that some of the systems may exceed the estimated galaxy halo escape velocities. The distribution of velocity separations between \mgii absorption and their host galaxies tends to be Gaussian with a mean offset of \absdeltavmean\ $\rm km\ s^{-1}$ and dispersion of about \absdeltavstd\ $\rm km\ s^{-1}$. In addition, we find that systems with equivalent width below the median, i.e. \ew $\le \medianmgiiew$\AA, show higher velocity dispersion with \absdeltavbelowmedianmgiiew\ $\rm km\ s^{-1}$, than the massive ones, hints that smaller clouds may leave the potential.

\section{Discussion and Summary}

Using the DECaLS imaging and SDSS fiber spectra, we search for the galaxies responsible for intervening \mgii absorbers over $0.4 < z < 1.0$. We detect \totalhostdetection\ galaxies hosting \totaldetection\ \mgii absorbers, based on the direct detection of nebular emission lines, at close impact parameters of $< 20$kpc around quasars with redshift ranging from \totaldetectionzabsmin\ to \totaldetectionzabsmax. Out of \totalhostdetection\ bonafide systems, the host galaxy's physical properties using multi-band SED fitting are derived for \primaryhostdetection\ galaxies (\primarydetection\ absorbers). Over the above redshift range, we find that the average detection fraction of \mgii absorber hosts, identified based on strong nebular emission, is 0.005 percent in fiber spectra. These systems preferentially belong to the strong \mgii absorbers with a median \ew \ of \medianmgiiew\AA, ranging between \mgiiewmin\AA\ $\le$ \ew $\le$ \mgiiewmax\AA. For ultra-strong \mgii absorbers with \ew $\ge 3$\AA, the detection rate of galaxies is factor 10 higher than the absorbers with \ew$\le 1$\AA. In contrast, in Paper I, we find an exceptionally high detection rate of 40\% for ultra-strong \mgii absorbers based on multi-band SED fitting of deep optical HSC-SSP Subaru and near-infrared VISTA imaging survey. The low detection rate in the present study can be explained by the fact that we only select the spectroscopically confirmed galaxies based on the direct detection of OII nebular emission, which is affected by fiber loss effects \citep{Lopez2012MNRAS.419.3553L},  as well as shallow imaging, and relatively poor seeing in DECaLS survey. In addition, the majority, i.e. \feiimgiigtzpfivepercentage\%, of systems show a strong \feii absorption with \ewfeii/\ew\ ratio $> 0.5$. It is in agreement with the result from \citet{Joshi2018MNRAS.476..210J} that indicates a strong correlation of \oii\ luminosity with \ew\ and $z$ of \mgii absorbers are mainly driven by such systems.

We find that the \ew\ strength as a function of $\rho$ is near constant at small impact parameters of $\rho \le 20$kpc, which implies a higher gas covering fraction.  \citet{Guha2024MNRAS.532.3056G} have found no anticorrelation between rest equivalent width of various ions, including Ca{\sc~ii}, Mn{\sc~ii}, Fe{\sc~ii}, Mg{\sc~ii}, and Mg{\sc~i}  absorption and impact parameter for 40 \mgii\ galaxies detected on top of quasars at  $0.37 \le z \le 1.01$ and at impact parameter range of 3-16 kpc.  In a high-resolution TNG50 cosmological magnetohydrodynamical simulation of the CGM around star-forming galaxies in $10^{11.5}-10^{12} M_{\odot}$ halos at $z \approx 1$, for strong \mgii absorber (\ew $ > 0.5$\AA) systems, \citet{DeFelippis2021ApJ...923...56D} found a constant \ew\ out to large impact parameters of $\sim 50 \rm kpc$.  For the \mgii systems with an impact parameter greater than $R_{90}$, largely tracing the disk-halo interface and galaxy halo, we see a strong anti-correlation between \ew\ and $\rho/R_{vir}$ suggesting that the \mgii gas is linked to the halo mass (see Figure~\ref{fig:EW_D}). \par

It is clear from Figure~\ref{fig:SFR_SM} that \mgii absorber host galaxies in our sample, as selected based on \oii\ nebular emission, follow the main sequence. These galaxies having a typical log(M$_{\star}$/M$_{\odot}$) of \stellarmassmin\ to \stellarmassmax\ and SFR of \sfgsfrmin\ to \sfgsfrmax\ $\rm M_{\odot}\rm yr^{-1}$, with a median SFR = \sfgsfrmedian\ $\rm M_{\odot}\ yr ^{-1}$. Interestingly, a handful of dwarf galaxies with log M$_{\star} < 9.5 \rm M_{\odot}$ in our sample show a starburst nature whereas \sfrpassivepercentage\% of galaxies show a passive nature with low average specific-SFR of $\rm log(sSFR) < -10.6 yr^{-1}$. Analyzing 381 star-forming galaxies between $0.70 < z < 2.34$ drawn from the MUSE Hubble Ultra Deep Survey \citet{Feltre2018A&A...617A..62F} have shown that galaxies hosting outflows, seen as blue-shifted \mgii absorption signature, preferentially show higher dust and neutral gas content in the interstellar medium. The average M$_{\star} = 10^{\stellarmassavg}\ M_{\odot}$ in our \mgii absorber hosts is consistent with the median stellar mass of $1.6 \times 10^{10}M_{\odot}$ of star-forming galaxies hosting outflows. 

For \mgii absorbers in our sample, typically having strong \ew\ of $\gtrsim 1$\AA, we find a mild increasing trend of \ew\ with the SFR (see Figure~\ref{fig:corr}). A similar trend is also evidenced with sSFR. Interestingly, considering only the systems tracing galaxy halo/disk-halo interface, with impact parameters greater than $R_{90}$, the \ew\ show a tighter correction with SFR and sSFR with Kendall's $\tau = 0.2$ ($p_{null} < 0.01$) and 0.14 ($p_{null} < 0.01$), respectively. In a galaxy probing galaxy experiment at a small impact parameter of $< 50$ kpc, \citet{Rubin2018ApJ...853...95R} found that galaxies with higher stellar mass and SFR have higher \ew. \citet{Lan2014ApJ...795...31L} analyzed the \mgii absorption properties between star-forming and passive galaxies defined by their colors and found an increasing \ew\ with M$_{\star}$, SFR and sSFR. In Paper-I, we also find a similar trend of \ew\ being correlated with the stellar mass and SFR for the ultra-strong \mgii\ systems \citep[see also,][]{Guha2023MNRAS.519.3319G}. Interestingly, no correction is observed between the \ew\ and stellar mass indicating that the positive correlation between \ew\ versus sSFR is driven by SFR and not M$_{\star}$. This is consistent with the observed trend of blue galaxies showing significantly stronger \ew\ \citep{Bordoloi2011ApJ...743...10B, Lan2014ApJ...795...31L}. On the other hand, in MUSE Analysis of Gas around Galaxies (MAGG) survey of 228 galaxies at $z \sim  0.8-1.5$ mapping large field scale environment with an average impact parameter of 165 kpc, \citet{Dutta2020MNRAS.499.5022D} have found that \ew\ strongly correlate with the stellar mass than SFR, but no correlation is seen with sSFR. The lack of correlation of \ew\ with M$_{\star}$ and a weak anti-correlation with the impact parameter in our sample can be attributed to the high gas covering fraction of about unity for the strong, \ew\ $> 1$\AA, systems at the small impact parameter ranges of $\le 50$ kpc as seen in simulations \citep{Ho2020ApJ...904...76H, nelson2020resolving, Ramesh:2023mox} and observations \citep{zahedy2019characterizing, Dutta2020MNRAS.499.5022D, anand2021characterizing}. 
Using 27 $z \approx 1$ host galaxies  with \ew $ > 0.5 - 0.8$\AA\  in  MusE GAs FLOw and Wind (MEGAFLOW) survey,  \citet{Schroetter2024A&A...687A..39S} have shown that the strong outflows show a tighter correlation with galaxy SFR and stellar mass \citep[see also,][]{Hopkins2012MNRAS.421.3522H}. Thus increasing \ew\ with SFR may hint at the origin of \mgii gas in galactic outflows.  \par
We finally find that the \mgii gas is largely bound to the dark matter halo of their host galaxies, except for a few systems.  The mean velocity difference of the absorbing gas and galaxy is \absdeltavmean\ $\rm km s^{-1}$, with a dispersion of \absdeltavstd\ $\rm km s^{-1}$. In a cold gas study using CGM-ZOOM scheme in AREPO, \citet{Suresh2019MNRAS.483.4040S} have shown that most gas is recycled more than once in galactic fountains. Using cosmological zoom-in simulation Eris2k \citet{Decataldo2024A&A...685A...8D} show that outflows are found to have a major contribution to the cold CGM gas budget at $z < 1$, with almost 50\% of the hot gas cooling in the outflow. Tracing this gas physics essentially demands probing the gas at close impact parameters of $\lesssim 50$ kpc,  where gas flows are more pronounced. A follow-up study of galaxies in this sample will enable us to probe the gas accretion and outflows near the disk-halo interface. 

\begin{acknowledgements}
LCH was supported by the National Science Foundation of China (11991052, 12233001), the National Key R\&D Program of China (2022YFF0503401), and the China Manned Space Project (CMS-CSST-2021-A04, CMS-CSST-2021-A06). 

Funding for the Sloan Digital Sky Survey IV has been provided by the Alfred P. Sloan Foundation, the U.S. Department of Energy Office of Science, and the Participating Institutions. 

SDSS-IV acknowledges support and resources from the Center for High Performance Computing at the University of Utah. The SDSS website is www.sdss4.org.

SDSS-IV is managed by the Astrophysical Research Consortium for the Participating Institutions of the SDSS Collaboration including the Brazilian Participation Group, the Carnegie Institution for Science, Carnegie Mellon University, Center for Astrophysics | Harvard \& Smithsonian, the Chilean Participation Group, the French Participation Group, Instituto de Astrof\'isica de Canarias, The Johns Hopkins University, Kavli Institute for the Physics and Mathematics of the Universe (IPMU) / University of Tokyo, the Korean Participation Group, Lawrence Berkeley National Laboratory, Leibniz Institut f\"ur Astrophysik Potsdam (AIP),  Max-Planck-Institut f\"ur Astronomie (MPIA Heidelberg), Max-Planck-Institut f\"ur Astrophysik (MPA Garching), Max-Planck-Institut f\"ur Extraterrestrische Physik (MPE), National Astronomical Observatories of China, New Mexico State University, New York University, University of Notre Dame, Observat\'ario Nacional / MCTI, The Ohio State University, Pennsylvania State University, Shanghai Astronomical Observatory, United Kingdom Participation Group, Universidad Nacional Aut\'onoma de M\'exico, University of Arizona, University of Colorado Boulder, University of Oxford, University of Portsmouth, University of Utah, University of Virginia, University of Washington, University of Wisconsin, Vanderbilt University, and Yale University.

The DESI Legacy Imaging Surveys consist of three individual and complementary projects: the Dark Energy Camera Legacy Survey (DECaLS), the Beijing-Arizona Sky Survey (BASS), and the Mayall z-band Legacy Survey (MzLS). DECaLS, BASS and MzLS together include data obtained, respectively, at the Blanco telescope, Cerro Tololo Inter-American Observatory, NSF’s NOIRLab; the Bok telescope, Steward Observatory, University of Arizona; and the Mayall telescope, Kitt Peak National Observatory, NOIRLab. NOIRLab is operated by the Association of Universities for Research in Astronomy (AURA) under a cooperative agreement with the National Science Foundation. Pipeline processing and analyses of the data were supported by NOIRLab and the Lawrence Berkeley National Laboratory (LBNL). Legacy Surveys also uses data products from the Near-Earth Object Wide-field Infrared Survey Explorer (NEOWISE), a project of the Jet Propulsion Laboratory/California Institute of Technology, funded by the National Aeronautics and Space Administration. Legacy Surveys was supported by: the Director, Office of Science, Office of High Energy Physics of the U.S. Department of Energy; the National Energy Research Scientific Computing Center, a DOE Office of Science User Facility; the U.S. National Science Foundation, Division of Astronomical Sciences; the National Astronomical Observatories of China, the Chinese Academy of Sciences and the Chinese National Natural Science Foundation. LBNL is managed by the Regents of the University of California under contract to the U.S. Department of Energy. The complete acknowledgments can be found at https://www.legacysurvey.org/acknowledgment/.

\end{acknowledgements}

%


%
%

\bibliographystyle{aa} 
\bibliography{aa} 
\end{document}

%% file: numbers.tex
\def\numqsosightlinesoverall{119,144}
\def\numuniqueabsorbersoverall{155,654}
\def\desicoveredqsosightlines{118,306}
\def\desicoveredabsorbers{154,596}

\def\desicoveredabsorberszmax{2.7}
\def\desitwoarcsecdetqso{8,565}
\def\desitwoarcsecdetabsorbers{11,037}
\def\desitwoarcsecdetapertures{8,792}
\def\desifinalsampleqso{4,812}
\def\desifinalsampleabsorbers{5,487}
\def\desifinalsampleapertures{5,508}
\def\oiitwopfivedetection{259}
\def\oiithreedetection{165}
\def\oiiihbadditionaldetection{11}
\def\totaldetection{270}
\def\totaldetectionzabsmin{0.4}
\def\totaldetectionzabsmax{1}
\def\vhsdetection{39}
\def\vikingdetection{7}
\def\primarydetection{213}
\def\secondarydetection{57}
\def\totalhostdetection{273}
\def\primaryhostdetection{216}
\def\threefilterdetection{251}
\def\threefilterdetectiongoodsed{216}
\def\noandbadsed{57}

\def\detectioniplowkpc{4}
\def\detectioniphighkpc{16}
\def\detectionzabsmin{0.4}
\def\detectionzabsmax{1}
\def\mgiiewgeonepercentage{91}
\def\medianmgiiew{2}
\def\mgiiewmin{0.27}
\def\mgiiewmax{4.72}
\def\feiimgiigtzpfivepercentage{84.26}
\def\feiimgiivsrhorvirstat{-0.09}
\def\feiimgiivsrhorvirpvalue{0.05}
\def\sfroiimin{0.18}
\def\sfroiimax{17.89}
\def\stellarmassmin{7.94}
\def\stellarmassmax{11.11}
\def\stellarmassavg{9.89}
\def\detectionsfg{194}

\def\sfgsfrmin{0.33}
\def\sfgsfrmax{218.27}
\def\sfgsfrmedian{5.81}
\def\passivesfrmean{0.15}

\def\sfrpassivepercentage{10.2}
\def\sfrstarburstpercentage{13.4}

\def\dwarfgalaxydetection{38}
\def\ebvmedian{0.14}
\def\massiveebvmean{0.25}
\def\massiveebverror{0.05}
\def\dwarfebvmean{0.12}
\def\dwarfebverror{0.04}
\def\absdeltavbelowmedianmgiiew{92}
\def\absdeltavmean{78}
\def\absdeltavstd{75}
\def\usmgiidetectionpercent{40}